\documentclass[12pt]{article}
\usepackage{amsmath}
\usepackage{prodint}
\usepackage{graphicx}
\usepackage{enumerate}
\usepackage{natbib}
\usepackage{url} 
\usepackage{binomexp}
\usepackage{mathtools}
\usepackage{amsfonts}
\usepackage{mathrsfs}
\RequirePackage{subcaption}
\usepackage{booktabs}
\usepackage{bbm}
\usepackage{dsfont}
\usepackage{longtable}
\usepackage{tabulary}
\usepackage{yhmath}
\usepackage{bm}
\usepackage{amssymb}
\usepackage[nodisplayskipstretch]{setspace}
\usepackage{tikz}
\usetikzlibrary{arrows.meta, decorations.pathmorphing, decorations.pathreplacing, positioning, calc, shadows.blur}
\usepackage{float}
\usepackage{color}
\usepackage{stmaryrd}
\usepackage{paralist}
\usepackage{multirow}
\usepackage{threeparttable}
\usepackage{soul}
\usepackage{algorithm,algpseudocode}
\makeatletter
\newcommand{\algmargin}{\the\ALG@thistlm}
\makeatother
\newlength{\whilewidth}
\algdef{SE}[parWHILE]{parWhile}{EndparWhile}[1]
  {\parbox[t]{\dimexpr\linewidth-\algmargin}{%
     \hangindent\whilewidth\strut\algorithmicwhile\ #1\ \algorithmicdo\strut}}{\algorithmicend\ \algorithmicwhile}%
\algnewcommand{\parState}[1]{\State%
  \parbox[t]{\dimexpr\linewidth-\algmargin}{\strut #1\strut}}

\DeclareFontFamily{U}{mathx}{\hyphenchar\font45}
\DeclareFontShape{U}{mathx}{m}{n}{<-6> mathx5 <6-7> mathx6 <7-8> mathx7 <8-9> mathx8 <9-10> mathx9 <10-12> mathx10 <12-> mathx12}{}
\DeclareSymbolFont{mathx}{U}{mathx}{m}{n}
\DeclareFontSubstitution{U}{mathx}{m}{n}
\DeclareMathAccent{\widecheck}{0}{mathx}{"71}

\newcommand{\bfs}{\bm{s}}

\newcommand{\bfy}{\bm{y}}

\newcommand{\bfdelta}{\bm{\delta}}
\newcommand{\bfeta}{\bm{\eta}}

\newcommand{\bftau}{\bm{\tau}}

\newcommand{\bfxi}{\bm{\xi}}

\newcommand{\rmd}{\mathrm{d}}

\newcommand{\bfS}{\bm{S}}

\newcommand{\bfW}{\bm{W}}

\newcommand{\bfY}{\bm{Y}}

\newcommand{\bbE}{\mathbb{E}}

\newcommand{\bbP}{\mathbb{P}}

\newcommand{\bbR}{\mathbb{R}}

\newcommand{\bbone}{\mathds{1}}

\newcommand{\calA}{\mathcal{A}}
\newcommand{\calB}{\mathcal{B}}
\newcommand{\calC}{\mathcal{C}}
\newcommand{\calD}{\mathcal{D}}

\newcommand{\calF}{\mathcal{F}}
\newcommand{\calH}{\mathcal{H}}
\newcommand{\calI}{\mathcal{I}}

\newcommand{\calS}{\mathcal{S}}

\usepackage[colorlinks=true,linkcolor=blue,citecolor=black]{hyperref}

\usepackage{amsthm}
\theoremstyle{plain}
\newtheorem{theorem}{Theorem}

\newtheorem{proposition}{Proposition}

\theoremstyle{definition}
\newtheorem{definition}{Definition}
\newtheorem{assumption}{Assumption}
\newtheorem{remark}{Remark}
\newtheorem{example}{Example}

\newcommand{\blind}{1}

\allowdisplaybreaks

\usetikzlibrary{shapes,decorations,arrows,calc,arrows.meta,fit,positioning}
\tikzset{
     -Latex,auto,node distance =1 cm and 1 cm,semithick,
     state/.style ={ellipse, draw, minimum width = 0.7 cm},
     point/.style = {circle, draw, inner sep=0.04cm,fill,node contents={}},
    bidirected/.style={Latex-Latex,dashed},
     el/.style = {inner sep=2pt, align=left, sloped}
 }

\begin{document}

\def\spacingset#1{\renewcommand{\baselinestretch}%
{#1}\small\normalsize} \spacingset{1}


\if1\blind
{
  \begin{center}
    {\Large \bf Randomization inference on cell effects under absorbing treatment onset}\\[0.5cm]
    {\large Xinyuan Chen}\\[0.2cm]
    Department of Mathematics and Statistics, Mississippi State University, \\Mississippi State, MS, USA\\
    xchen@math.msstate.edu
  \end{center}
} \fi

\if0\blind
{
  \bigskip
  \bigskip
  \bigskip
  \begin{center}
    {\Large \bf Randomization inference on cell effects under absorbing treatment onset}
\end{center}
  \medskip
} \fi

\begin{abstract}
    In multiple-baseline experiments, staggered adoption designs, and stepped-wedge trials, every unit switches once from baseline to treatment at a randomized onset time and remains treated afterward. The treatment onsets in these designs are absorbing. Since such experiments often involve few heterogeneous units, randomization tests of the sharp null of no effect on any unit-period cell are common. Rejecting this null, however, implies that some effect exists somewhere, not that it is large, positive for most unit-periods, or persistent. To improve on this limitation, we study what can be established via randomization tests for bounded and quantile nulls on cell effects in absorbing onset designs. Under the assumption that a cell's potential outcome depends on its current treatment status and not on when treatment began, we establish finite-sample validity for both tests. We then develop the power theory for the bounded-null tests with a fixed number of units $N$ and a growing series length $T$ admitting $K_T$ onsets. The $p$-value against a fixed alternative decays as $K_T^{-N}$ up to a factor polylogarithmic in $T$, with lower bounds of the same order, so the onset window, not the series length, determines the power. Theory-implied guidelines for experiment planning are also provided.
\end{abstract}

\noindent%

\spacingset{1.75} 

\section{Introduction}\label{sec:intro}

In a staggered rollout, every unit begins untreated, then switches to treatment at some period, and stays treated afterward. Therefore, a unit's entire treatment path is determined by its onset time, and the treatment onset is absorbing. Such structures can be found in multiple-baseline experiments in single-case research \citep[e.g.,][]{Edgington2007, Kratochwill2010, Onghena2024}, staggered adoption designs \citep[e.g.,][]{Athey2022, Roth2023, Xiong2024}, and stepped-wedge cluster trials \citep[e.g.,][]{Hussey2007, Ji2017}. In a design-based framework, the onset times are randomized, and potential outcomes and covariates are treated as fixed. 

One feature in staggered rollouts is that the units are often few and heterogeneous. For example, in a review of 160 stepped-wedge trials published between 2016 and 2022, the median trial randomized 11 clusters to five sequences \citep{Nevins2024}. 61 trials from the same period had between two and six clusters \citep{Tong2025}, and the modal single-case study uses a multiple-baseline design with three or four cases \citep{Shadish2011}. Hence, the finite-sample randomization tests are often adopted in these analyses for the sharp null hypothesis that the treatment has no effect on any unit at any period \citep[e.g.,][]{Edgington2007, Levin2021, Michiels2019, Wang2017, KennedyShaffer2020}. Rejecting this sharp null, however, implies that some effect exists somewhere, not that it is large, positive for most unit-periods, or persistent. 

Recently, new methods have been developed to address this limitation, e.g., the interval method of \citet{Michiels2017}, the conditional randomization tests of \citet{Shaikh2021} and \citet{Zhang2025}, and the permutation test of \citet{Roth2023}. A related stream in econometrics treats few treated units and targets the average effect on them \citep[e.g.,][]{Conley2011, Ferman2019, Alvarez2025, Hagemann2025} or tests a fully specified trajectory of effects for a treated unit \citep{Chernozhukov2021}. Each of these methods, however, either describes the effect with a single number, e.g., a constant shift common to every cell or an average over the cells, or specifies every effect exactly. To fully address the issues, the object of interest should be shifted to the matrix of cell effects, $\bftau=(\tau_{it})$, where $\tau_{it}$ is the effect of unit $i$ at period $t$, one number per unit-period cell $(i,t)$ (properly defined under certain structural assumptions). The natural questions are whether some $\tau_{it}$ exceeds a given amount and whether $\tau_{it}$ exceeds a threshold in a stated fraction of the cells. The null hypotheses for these questions are composite. In particular, a \emph{bounded null} states that every $\tau_{it}$ is at most a prescribed bound $\delta_{it}$, so that rejecting it establishes that some cell has an effect above its bound. A \emph{quantile null} states that the $k$th smallest of the effects in a set of $B$ cells is at most $c$, so that rejecting it establishes that at least $B-k+1$ of these cells have effects above $c$. Although cell effects are not point identified because each cell reveals only one potential outcome, these hypotheses on them are testable. Specifically, the null confines the unobserved potential outcomes to a set of tables (examples given in Section S1 of the Supplement), the null-compatible set, and a test is valid if it controls the level $\alpha$ at every table in the set.

The literature in this area has focused on single-period experiments, including completely randomized, stratified, and matched designs \citep[e.g.,][]{Caughey2023, Su2024, ChenRothSpiess2025, Chen2025, Kim2026}, and on single-period experiments with interference \citep{Huang2025}. Absorbing onset designs change the problem in three notable ways. First, the unit of inference is a unit-period cell, and a counterfactual onset flips a contiguous block of a unit's cells and leaves the rest fixed. Consequently, which cells an unbounded effect can influence depends on the design and on where the cell lies relative to the realized onset, not on the observed outcomes alone. Because the onsets move only contiguous blocks of cells, the reference distribution of a rank statistic also depends on the outcomes. The closed-form maximizer of the single-period case \citep{Caughey2023} then need not attain the supremum of the $p$-value. Second, a cell can be treated under both the realized and a counterfactual onset while the two onsets differ, so imputing the counterfactual requires the potential outcome of a treated cell under a different treatment history, which a single-period design does not possess. Third, each unit is randomized once, over its onset time, so the reference distribution grows with the number of admissible onsets per unit and the number of units, not directly with the series length. 

In this paper, we study what randomization tests can establish for the bounded and quantile nulls on cell effects under absorbing onset designs. Our findings are as follows. First, under the treatment-duration irrelevance assumption (Assumption \ref{ass:static}), a cell's outcome depends on when treatment began only through its current status. Each unit-period cell can then be viewed as a unit with a binary treatment, and the finite-sample validity result of \citet{Caughey2023} for effect-increasing statistics holds for the bounded-null test under any onset design. Second, for quantile nulls, the supremum of the bounded-null $p$-value over the effect tables the null allows is a valid $p$-value under every onset design. We reduce it to a maximum over finitely many cell sets and, for the Wilcoxon rank statistic, bound it by an envelope whose terms for each unit have a closed form. Third, with a fixed number of units $N$ and a growing number $K_T$ of admissible onsets, the $p$-value against a fixed alternative decays as $K_T^{-N}$, and, for sub-Gaussian baselines, the series length enters only through a factor polylogarithmic in $T$. Lower bounds of the same order show that the onset window, not the series length, determines the power. Section \ref{sec:planning} turns these results into rules for planning an experiment and calibrates by simulation the window needed for a target power. Last but not least, Section S4 of the Supplement treats settings in which Assumption \ref{ass:static} fails. If only one part of it is imposed, the supremum of the mean-contrast randomization $p$-value over the tables that agree with the data and satisfy a one-sided bounded null equals $1$ for every data set. A null that also bounds how much a cell's outcome may change between onsets gives a valid test and a sensitivity analysis with one parameter for duration dependence and one for anticipation.

The rest of the article is organized as follows. Section \ref{sec:framework} sets up the framework. Section \ref{sec:bounded} treats bounded nulls, Section \ref{sec:quantiles} quantile nulls, and Section \ref{sec:power} the power theory. Section \ref{sec:planning} turns the power theory into rules for planning an experiment, and Section \ref{sec:data} presents an application to a multiple-baseline experiment. Section S1 of the Supplement works a small example and treats cluster randomized designs. Sections S2 and S3 give the proofs, the worst-case rates over the baseline classes, and the regulated designs. Section S4 treats onset-dependent potential outcomes and the sensitivity analysis, Section S5 reports the simulation studies, and Section S6 gives additional details and analyses of the application.

\section{Framework}\label{sec:framework}

\subsection{Absorbing onset}\label{sec:onset}

Consider a randomized experiment with $N$ units observed over periods $t\in\{1,\ldots,T\}$. Unit $i\in\{1,\ldots,N\}$ begins untreated, switches to treatment at an onset time $S_i$, and remains treated thereafter. The admissible onset times comprise a set $\calS\subseteq\{2,\ldots,T\}$. Period $1$ is excluded from $\calS$ so that each unit has at least one pre-onset time. Every unit switches status during the study, and designs with a never-treated group are not covered. The onset vector $\bfS=(S_1,\ldots,S_N)$ is drawn from a known design distribution $\bbP$ on $\calS^N$ with support $\Omega$, which is the sole source of randomness. We write $W_{it}(s)=\bbone(t\ge s)$ as the treatment status of cell $(i,t)$ when unit $i$ has onset $s$, and $\bfW(\bfs)$ as the $N\times T$ status matrix induced by $\bfs\in\Omega$. (Table S1 of the Supplement illustrates.) Throughout the main text, we use $\bbP$ and $\bbE$ for probabilities and expectations over the design, and $\bbP_{\bfs'}$ for the probability over an independent copy $\bfs'$ of $\bfS$ with everything else held fixed. Example \ref{ex:designs} gives some commonly used onset designs.

\begin{example}[Onset designs]\label{ex:designs}
    (a) \emph{Independent onsets} \citep{Marascuilo1988}. $S_1,\ldots,S_N$ are independent, with $S_i$ being uniform over a window $\calS=\{a+1,\ldots,a+K\}$ of $K$ consecutive periods starting after period $a$, so that $|\Omega|=K^N$. (b) \emph{Permuted onsets} \citep{Wampold1986}. $N$ distinct onset times $s_1<\cdots<s_N$ are fixed in advance and assigned to the units by a uniform random permutation, so that $|\Omega|=N!$. (c) \emph{Restricted onsets.} Any design supported on a subset of $\calS^N$, e.g., the regulated randomization of \citet{Koehler1998}, which assigns the units at random to staggered windows and draws an onset within each, the restricted scheme of \citet{Levin2018}, which imposes a minimum spacing between onsets, or a stepped-wedge trial that randomly partitions clusters into sequences, which is of type (b) when exactly one cluster switches at each step.
\end{example}

\begin{figure}[t]
    \centering
    \includegraphics[width=\textwidth]{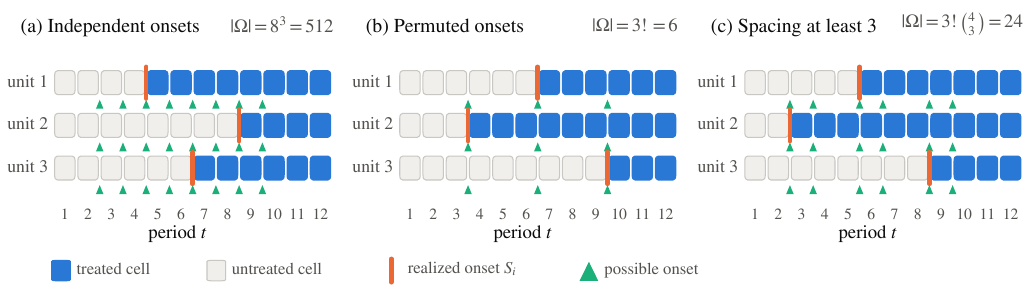}
    \caption{The onset designs in Example \ref{ex:designs} with $N=3$ units and $T=12$ periods. Each panel shows one realized onset vector. Blue cells are treated, gray cells are untreated, the orange bar marks the realized onset $S_i$, and the green triangles mark the onsets each unit could have received. (a) Independent onsets uniform on $\{3,\ldots,10\}$, realized $(5,9,7)$. (b) Permuted onsets, with the fixed onsets $4$, $7$, $10$ assigned by a uniform permutation, realized $(7,4,10)$. (c) Independent onsets on $\{3,\ldots,10\}$ conditioned on a spacing of at least $3$ between every pair, realized $(6,3,9)$. This design has $24$ atoms.}\label{fig:designs}
\end{figure}

Figure \ref{fig:designs} illustrates the three designs in Example \ref{ex:designs}. In practice, units may be observed over different series lengths $T_i$, with admissible onsets $\calS_i\subseteq\{2,\ldots,T_i\}$. We write $T$ and $\calS$ for notational simplicity, since every statement in Sections \ref{sec:bounded} and \ref{sec:quantiles}, and in Section S4 of the Supplement, holds with $(T_i,\calS_i)$ in place of $(T,\calS)$. 

\subsection{Potential outcomes}\label{sec:po}

The potential outcomes of cell $(i,t)$ are indexed by the onset, $Y_{it}(s)$ for $s\in\calS$, since the treatment path of a unit is a function of its onset. This is the path-indexed formulation of \citet{Bojinov2019} and \citet{Bojinov2021}, restricted to absorbing paths, and that of \citet{Athey2022} for staggered adoption. Under the stable unit treatment value assumption (SUTVA), units do not interfere, so $Y_{it}(\cdot)$ depends on $\bfs=(s_1,\ldots,s_N)$ only through $s_i$. Each cell admits $|\calS|$ potential outcomes, and we observe $Y_{it}=Y_{it}(S_i)$, with the realized $N\times T$ matrix being $\bfY=\bfY(\bfS)$. 

\begin{assumption}[Treatment-duration irrelevance]\label{ass:static}
    For every cell $(i,t)$ and all $s,s'\in\calS$: (i) if $\max(s,s')\le t$, then $Y_{it}(s)=Y_{it}(s')$; (ii) if $\min(s,s')>t$, then $Y_{it}(s)=Y_{it}(s')$.
\end{assumption}

Assumption \ref{ass:static}(i) implies that the outcome for a treated cell is unaffected by the duration of treatment, while Assumption \ref{ass:static}(ii) implies that the outcome for an untreated cell is unaffected by the timing of treatment onset. Together they assume that $Y_{it}(s)$ depends on $s$ only through the status $W_{it}(s)$, so that each cell is reduced to having two potential outcomes, $Y_{it}^{(1)}$ and $Y_{it}^{(0)}$, with $Y_{it}(s)=Y_{it}^{(1)}W_{it}(s)+Y_{it}^{(0)}\{1-W_{it}(s)\}$ and $\tau_{it}=Y_{it}^{(1)}-Y_{it}^{(0)}$. The effect matrix $\bftau=(\tau_{it})$ has $N\times T$ entries. For cells treated under all admissible onsets ($t\ge\max\calS$), or under none ($t<\min\calS$), no onset reveals one of the two potential outcomes. This outcome is an unknown constant of the table, constrained only through $\tau_{it}$ (Section S1 of the Supplement gives an example).

Assumption \ref{ass:static} is the treatment-duration irrelevance assumption of \citet{ChenLi2025} for stepped-wedge designs. Part (i) is the invariance-to-history assumption of \citet{Athey2022}, and part (ii) is the no-anticipation assumption of \citet{Athey2022} and the nonanticipating potential outcomes of \citet[Assumption 1]{Bojinov2021}. The sharp null of no effect, $Y_{it}(s)=Y_{it}(s')$ for all $s,s'\in\calS$, implies Assumption \ref{ass:static} and sets every effect to zero. The bounded and quantile nulls of Section \ref{sec:nulls} keep Assumption \ref{ass:static} and allow nonzero effects.

\subsection{Null hypotheses}\label{sec:nulls}

Under Assumption \ref{ass:static}, let $\bfdelta=(\delta_{it})$ be an $N\times T$ matrix of bounds with entries in $\bbR_{+\infty}\equiv\bbR\cup\{+\infty\}$. The sharp null, for finite $\bfdelta$, and the bounded null are
\begin{equation*}
    H_{\bfdelta}:\ \tau_{it}=\delta_{it}\ \text{for all }(i,t),\quad\text{and}\quad H_{\le\bfdelta}:\ \tau_{it}\le\delta_{it}\ \text{for all }(i,t),
\end{equation*}
where a bound $\delta_{it}=+\infty$ in $H_{\le\bfdelta}$ makes $\tau_{it}$ unrestricted. For a set of cells $\calB\subseteq\{1,\ldots,N\}\times\{1,\ldots,T\}$ with $|\calB|=B$, let $\tau_{(1)}^{\calB}\le\cdots\le\tau_{(B)}^{\calB}$ be the order statistics of $\{\tau_{it}:(i,t)\in\calB\}$. Then, the quantile null is
\begin{equation}\label{eq:quantile-null}
    H_{k,c}^{\calB}:\ \tau_{(k)}^{\calB}\le c,
\end{equation}
for $k\in\{1,\ldots,B\}$ and $c\in\bbR$, and we write $\calH_{k,c}^{\calB}=\{\bftau\in\bbR^{N\times T}:\tau_{(k)}^{\calB}\le c\}$ as the set of effect matrices it contains. Rejecting $H_{k,c}^{\calB}$ means that at least $B-k+1$ of the cells in $\calB$ have effects exceeding $c$. These quantiles are order statistics of the cell effects, not differences of marginal quantiles. Natural choices of $\calB$ are all $NT$ cells and the $T$ cells of a single unit. Nulls with the inequalities reversed, such as $H_{\ge\bfdelta}:\tau_{it}\ge\delta_{it}$ for all $(i,t)$, are tested by applying the tests to $-\bfY$ with the bound $-\bfdelta$. For quantile nulls, the threshold becomes $-c$ and the index $B-k+1$. A two-sided null such as $H_{\le\bfdelta+\epsilon}\cap H_{\ge\bfdelta-\epsilon}$, with a constant $\epsilon>0$, is rejected at level $\alpha$ when either one-sided test rejects at level $\alpha/2$. The statement that at least $B-k+1$ cells have effects above $c$ is an attributable effect in the sense of \citet{Rosenbaum2001, Rosenbaum2007}.

\subsection{Imputation and the randomization $p$-value}\label{sec:imputation}

Consider $\bfdelta$ fixed. Given $(\bfS,\bfY)$ and finite $\bfdelta$, the potential-outcome table imputed under $H_{\bfdelta}$ is composed of $\widetilde Y_{it}^{(0)}=Y_{it}-\delta_{it}W_{it}(S_i)$ and $\widetilde Y_{it}^{(1)}=Y_{it}+\delta_{it}\{1-W_{it}(S_i)\}$, and the imputed observed matrix under a counterfactual onset vector $\bfs'$ is $\widetilde\bfY(\bfs')$ with entries
\begin{equation}\label{eq:impute-cf}
    \widetilde Y_{it}(\bfs')=\widetilde Y_{it}^{(1)}W_{it}(s_i')+\widetilde Y_{it}^{(0)}\{1-W_{it}(s_i')\}=\begin{cases}Y_{it}+\delta_{it},& W_{it}(s_i')=1,\ W_{it}(S_i)=0,\\ Y_{it}-\delta_{it},& W_{it}(s_i')=0,\ W_{it}(S_i)=1,\\ Y_{it},&\text{otherwise}.\end{cases}
\end{equation}
For $\delta_{it}=+\infty$, the imputed matrix is defined by the three-case rule on the right of \eqref{eq:impute-cf}, with $Y_{it}\pm\infty=\pm\infty$. We call a cell \emph{flipped} under $(S_i,s_i')$ if its treatment status differs under the two onsets, i.e., $W_{it}(S_i)\ne W_{it}(s_i')$. The flipped cells of unit $i$ are at the periods between $\min(S_i,s_i')$ and $\max(S_i,s_i')-1$. Note that $\widetilde\bfY(\bfS)=\bfY$. For a test statistic $\psi:\Omega\times\overline{\bbR}^{N\times T}\to\overline{\bbR}$, where $\overline{\bbR}\equiv\bbR\cup\{\pm\infty\}$, the randomization $p$-value is
\begin{equation}\label{eq:pvalue}
    p_{\bfS,\bfdelta}=\bbP_{\bfs'}\left[\psi\{\bfs',\widetilde\bfY(\bfs')\}\ge\psi(\bfS,\bfY)\right].
\end{equation}
Under $H_{\bfdelta}$ with finite $\bfdelta$, the imputed table is the true table, and $p_{\bfS,\bfdelta}$ is the Fisher randomization $p$-value \citep{Fisher1935}, valid in the sense that $\bbP(p_{\bfS,\bfdelta}\le\alpha)\le\alpha$ for every $\alpha\in[0,1]$. Table S3 of the Supplement gives an example.

\begin{definition}[Effect-increasing statistic]\label{def:ei}
    For $\bfs\in\Omega$, let $\calD(\bfs)$ be the set of matrices $\bfy\in\overline{\bbR}^{N\times T}$ with $y_{it}=+\infty$ only if $W_{it}(s_i)=1$ and $y_{it}=-\infty$ only if $W_{it}(s_i)=0$. The statistic $\psi$ is effect-increasing if for every $\bfs\in\Omega$, $\bfy\in\calD(\bfs)$, and $\bfeta,\bfxi\in\overline{\bbR}^{N\times T}$ with $\bfeta\ge0\ge\bfxi$ entrywise,
    \begin{equation}\label{eq:ei}
        \psi\left[\bfs,\ \bfy+\bfW(\bfs)\circ\bfeta+\{1-\bfW(\bfs)\}\circ\bfxi\right]\ \ge\ \psi(\bfs,\bfy),
    \end{equation}
    where $\circ$ is the entrywise product, and sums are in the extended reals with the conventions that $\pm\infty$ plus a finite number is $\pm\infty$, $(\pm\infty)+(\pm\infty)=\pm\infty$, and $0\cdot(\pm\infty)=0$. The perturbed matrix then lies in $\calD(\bfs)$, and no entry is of the form $(+\infty)+(-\infty)$.
\end{definition}

Definition \ref{def:ei} is the cell-wise version of the effect-increasing property in \citet{Rosenbaum2002} and \citet{Caughey2023}, such that raising the outcome of a cell treated under $\bfs$, or lowering the outcome of an untreated cell under $\bfs$, does not decrease the statistic. Examples \ref{ex:stats} and \ref{ex:ranks} give the families of $\psi$ used below.

\begin{example}[Mean contrasts and linear statistics]\label{ex:stats}
    (a) \emph{Mean contrasts.} For unit $i$ with onset $s\in\calS$ and outcome row $\bfy_i=(y_{i1},\ldots,y_{iT})$, let $\gamma_i(s;\bfy_i)=(T-s+1)^{-1}\sum_{t\ge s}y_{it}-(s-1)^{-1}\sum_{t<s}y_{it}$, i.e., the post-onset mean minus the pre-onset mean, both blocks being nonempty because $2\le s\le T$. Then, $\psi_M(\bfs,\bfy)=\sum_{i=1}^N\gamma_i(s_i;\bfy_i)$, and any weighted sum of the $\gamma_i$ with nonnegative weights is effect-increasing, since $\gamma_i(s;\cdot)$ is nondecreasing in every post-onset cell and nonincreasing in every pre-onset cell. (b) \emph{Linear statistics.} Let $w^{+}_{it},w^{-}_{it}\ge0$ be fixed weights and $\psi_L(\bfs,\bfy)=\sum_{i,t}[W_{it}(s_i)w^{+}_{it}-\{1-W_{it}(s_i)\}w^{-}_{it}]y_{it}$, i.e., a weighted sum of the treated cells minus a weighted sum of the untreated cells, the weight of a cell depending on its status under $\bfs$, not on the onset itself. $\psi_L(\bfs,\bfy)$ is effect-increasing because the coefficient of $y_{it}$ is $w^{+}_{it}\ge0$ at a cell treated under $\bfs$ and $-w^{-}_{it}\le0$ at an untreated cell. The same holds for onset-dependent weights $w^{+}_{it}(s_i),w^{-}_{it}(s_i)\ge0$, of which $\psi_M$ is the instance $w^{+}_{it}(s)=(T-s+1)^{-1}$, $w^{-}_{it}(s)=(s-1)^{-1}$.
\end{example}

\begin{example}[Within-unit rank sums]\label{ex:ranks}
    Let $R_{it}(\bfy_i)$ be the rank of $y_{it}$ among $y_{i1},\ldots,y_{iT}$, with ties broken by period index, the earlier period receiving the lower rank. Let $\phi_i$ be nondecreasing and nonconstant scores, and let $\Psi_i(s;\bfy_i)=\sum_{t=1}^TW_{it}(s)\phi_i\{R_{it}(\bfy_i)\}$ be the score sum of the post-onset cells. Its distribution over a uniformly random set of $T-s+1$ ranks has mean $m_i(s)=(T-s+1)\overline\phi_i$ and variance $v_i^2(s)=(T-s+1)(s-1)(T-1)^{-1}T^{-1}\sum_{r=1}^T\{\phi_i(r)-\overline\phi_i\}^2$, with $\overline\phi_i$ the average score. Here $v_i(s)>0$ for every $s\in\{2,\ldots,T\}$ because both blocks are nonempty and the scores are nonconstant. For the Wilcoxon scores, $\phi_i(r)=r$, $m_i(s)=(T-s+1)(T+1)/2$ and $v_i^2(s)=(T-s+1)(s-1)(T+1)/12$. The standardized within-unit rank sum is $\psi_R(\bfs,\bfy)=\sum_{i=1}^N\{\Psi_i(s_i;\bfy_i)-m_i(s_i)\}/v_i(s_i)$, the standardization making the summands comparable across onsets, whose post-onset blocks differ in size. $\psi_R(\bfs,\bfy)$ is effect-increasing, because for fixed $s$ each summand is an increasing affine function of $\Psi_i(s;\bfy_i)$, which itself is effect-increasing (Section S2.1 of the Supplement). For a bound matrix $\bfdelta$ with finite entries, the $\bfdelta$-adjusted rank statistic is $\psi_R^{\bfdelta}(\bfs,\bfy)=\psi_R\{\bfs,\bfy-\bfdelta\circ\bfW(\bfs)\}$, the rank statistic applied to the matrix in which the bound has been subtracted from every cell that is treated under $\bfs$. $\psi_R^{\bfdelta}(\bfs,\bfy)$ is also effect-increasing because the subtraction depends on a cell only through its status under $\bfs$, so a perturbation of the form \eqref{eq:ei} for $\bfy$ is a perturbation of the same form for $\bfy-\bfdelta\circ\bfW(\bfs)$. Table S4 of the Supplement illustrates $\psi_R(\bfs,\bfy)$ and $\psi_R^{\bfdelta}(\bfs,\bfy)$.
\end{example}

The mean contrast is the statistic of the classical multiple-baseline randomization test \citep[e.g.,][]{Wampold1986, Levin2021}. Both families are sums of unit summands, and each summand depends only on that unit's row. In a stepped-wedge cluster randomized trial, the unit is a cluster, and the cell outcome is a weighted average of the individual outcomes in the cluster-period, or alternatively, the individual-period itself can be viewed as the cell. Remark S1 of the Supplement gives the details, including the block sizes for the rank statistic and the condition under which the power theory of Section \ref{sec:power} applies.

\section{Bounded nulls}\label{sec:bounded}

\begin{theorem}[Validity for bounded nulls]\label{thm:bounded}
    Let Assumption \ref{ass:static} hold and $\psi$ be effect-increasing. Then, for every design $\bbP$ on $\calS^N$ and every $\bfdelta\in\bbR_{+\infty}^{N\times T}$, the $p$-value $p_{\bfS,\bfdelta}$ in \eqref{eq:pvalue} satisfies $\bbP(p_{\bfS,\bfdelta}\le\alpha)\le\alpha$ for all $\alpha\in[0,1]$ under $H_{\le\bfdelta}$.
\end{theorem}

The proof is in Section S2.1 of the Supplement, and it covers bounds equal to $+\infty$. Under $H_{\le\bfdelta}$, the imputed matrix $\widetilde\bfY(\bfs')$ is at least the true matrix at every cell treated under $\bfs'$ and at most the true matrix at every untreated cell. An effect-increasing statistic is therefore at least as large in the imputed matrix at each $\bfs'$. The test can be conservative when $\tau_{it}<\delta_{it}$ at cells that some counterfactual onset flips. At the boundary $H_{\bfdelta}$, the $p$-value is the Fisher randomization $p$-value, and the level is attained up to the discreteness of the reference distribution. Section S5.1 of the Supplement checks the size of the test via simulation. Taking $\bfdelta=\bm0$ recovers the classical multiple-baseline randomization test, whose $p$-value is thus also valid for the broader hypothesis that no cell has a positive effect. Taking $\delta_{it}=\delta$, a constant, and collecting the values of $\delta$ under which $H_{\le\bfdelta}$ is not rejected gives a one-sided lower confidence bound for $\max_{i,t}\tau_{it}$ that requires no homogeneity across units or periods, in contrast to the constant-effect inversion of \citet{Michiels2017}.

\begin{proposition}[Monotonicity in the bound]\label{prop:mono}
    Let $\psi$ be effect-increasing. Then, for every design and every realized $(\bfS,\bfY)$, the map $\bfdelta\mapsto p_{\bfS,\bfdelta}$ is nondecreasing in the entrywise order on $\bbR_{+\infty}^{N\times T}$.
\end{proposition}

Proposition \ref{prop:mono} states that the $p$-value of the bounded null is monotone in the bound. The proof, in Section S2.2 of the Supplement, applies Definition \ref{def:ei} to the difference between two imputations, onset by onset.

\begin{remark}[Relation to the single-period theory]\label{rem:caughey-bounded}
    For finite bounds, Theorem \ref{thm:bounded} is Theorem 1(a) of \citet{Caughey2023} applied to the $NT$ cells, with the onset design as the assignment mechanism, allowed by Assumption \ref{ass:static}. Under it, each cell has two potential outcomes, and the bounded null determines one imputed table that bounds the $p$-value of every table in the null. If only one part of Assumption \ref{ass:static} is imposed, the outcome of a cell in one of the two statuses can differ across onsets, and a bounded null on the onset-indexed effects no longer determines such a table. Theorem S3 of the Supplement shows that the supremum of the randomization $p$-value of the mean contrast over the tables that agree with the data and satisfy such a null then equals 1 for every data set. Section S4.2 of the Supplement instead bounds how much a cell's outcome may change between any two onsets, which restores a valid test and gives a sensitivity analysis.
\end{remark}

\section{Quantiles of cell effects}\label{sec:quantiles}

\subsection{The supremum over the null}\label{sec:sup}

Consider a fixed set of units $\calI\subseteq\{1,\ldots,N\}$. Let $\calB=\{(i,t):i\in \calI,\ 1\le t\le T\}$ be the union of their rows, with $|\calB|=B=|\calI|T$, and fix an index $k$ and a constant $c$. Throughout this section, the statistic $\psi$ is assumed to depend on $\bfy$ only through the rows in $\calB$. For the statistics in Examples \ref{ex:stats} and \ref{ex:ranks}, this means that $\psi$ is the sum of the unit summands over $i\in \calI$, which for $\calI=\{1,\ldots,N\}$ is the full statistic. Under this premise, the bounds $\delta_{it}$ at cells outside $\calB$ do not affect $p_{\bfS,\bfdelta}$ and can be left unspecified. The randomization $p$-value of the quantile null $H_{k,c}^{\calB}$ in \eqref{eq:quantile-null} is the supremum of the bounded-null $p$-values over the bound matrices the null allows, i.e.,
\begin{equation}\label{eq:sup-p}
    p_{\bfS,k,c}=\sup\left\{p_{\bfS,\bfdelta}:\ \bfdelta\in\bbR_{+\infty}^{N\times T},\ \delta_{(k)}^{\calB}\le c\right\},
\end{equation}
\vspace{-0.2cm}
where $\delta_{(k)}^{\calB}$ is the $k$th smallest of the entries of $\bfdelta$ on $\calB$ and infinite bounds act on the imputation as in \eqref{eq:impute-cf}. The supremum is attained on a finite set of bound matrices. Specifically, for a set of cells $\calA\subseteq\calB$ with $|\calA|=B-k$, let $\bfdelta(\calA)$ be the bound matrix with $\delta_{it}(\calA)=+\infty$ for $(i,t)\in\calA$ and $\delta_{it}(\calA)=c$ for $(i,t)\in\calB\setminus\calA$. It leaves the $B-k$ cells of $\calA$ unbounded and bounds the other $k$ cells of $\calB$ by $c$, so $H_{\le\bfdelta(\calA)}$ is contained in $H_{k,c}^{\calB}$.

\begin{theorem}[Quantile nulls]\label{thm:quantile}
    Let Assumption \ref{ass:static} hold and $\psi$ be effect-increasing and depend on $\bfy$ only through the rows in $\calB$. Then, for every design, $\bbP(p_{\bfS,k,c}\le\alpha)\le\alpha$ for all $\alpha$ under $H_{k,c}^{\calB}$, and
    \begin{equation}\label{eq:max-A}
        p_{\bfS,k,c}=\max_{\calA\subseteq\calB,\ |\calA|=B-k}\ p_{\bfS,\bfdelta(\calA)}.
    \end{equation}
\end{theorem}

The proof is in Section S2.3 of the Supplement. Under the imputation in \eqref{eq:impute-cf}, a cell $(i,t)\in\calA$ affects the reference statistic at $\bfs'$ only if $\bfs'$ flips it, i.e., a treated cell at $t\ge S_i$ is flipped by the onsets $s_i'>t$, and an untreated cell at $t<S_i$ by the onsets $s_i'\le t$. The maximizing set can therefore change with $\bfs'$. In the example of Section S1 of the Supplement, it is the cell at the realized onset, not the cell with the largest observed value.

A cell $(i,t)$ with $t<\min\calS$ is untreated under every admissible onset, and a cell with $t\ge\max\calS$ is treated under every one. No counterfactual onset flips such a cell, so its bound does not affect the $p$-value. Let $\calF$ be the set of cells $(i,t)$ with $\min\calS\le t<\max\calS$. The test of $H_{k,c}^{\calB}$ is then valid for the larger null that at most $B-k$ cells of $\calB\cap\calF$ have effects above $c$. A rejection thus shows that at least $B-k+1$ cells of $\calB\cap\calF$ have effects above $c$. If $B-k\ge|\calB\cap\calF|$, this larger null holds for every effect matrix, and the test rejects with probability at most $\alpha$ whatever the effects. Section S2.3 of the Supplement proves these statements.

Two more consequences follow from which cells are flipped. First, the two cells next to the onset, $(i,S_i)$ and $(i,S_i-1)$, are flipped by the largest set of counterfactual onsets, i.e., $(i,S_i)$ by every $s_i'>S_i$, and $(i,S_i-1)$ by every $s_i'<S_i$. This leads to \eqref{eq:linear-burn} below. Under a statistic with a bounded response to a single cell, they need not be the maximizing cells. Second, a statistic that responds to a single infinite cell by an infinite amount gives a quantile test with little power. Under $\psi_L$ with $w^{+}_{it},w^{-}_{it}>0$ (or $\psi_M$), the cell $(i,S_i)$ at $-\infty$ sends the reference statistic to $+\infty$ at every $\bfs'$ with $s_i'>S_i$, and the cell $(i,S_i-1)$ at $+\infty$ does the same at every $\bfs'$ with $s_i'<S_i$, so for $B-k\ge1$, we have
\begin{equation}\label{eq:linear-burn}
    p_{\bfS,k,c}\ \ge\ \max_{i\in \calI}\max\left\{\bbP_{\bfs'}(s_i'>S_i),\ \bbP_{\bfs'}(s_i'<S_i)\right\}\ \ge\ \frac12\left\{1-\min_{i\in \calI}\bbP_{\bfs'}(s_i'=S_i)\right\},
\end{equation}
which approaches $1/2$ as the marginal probabilities $\bbP_{\bfs'}(s_i'=S_i)$ become small. Under the designs in Example \ref{ex:designs}(a) with $K\ge2$, the bound is at least $1/4$, so the quantile test with these statistics cannot reject at level $\alpha<1/4$. On the other hand, under $\psi_R$ with scores $\phi_i$, a cell at $\pm\infty$ affects $\Psi_i$ by at most $\phi_i(T)-\phi_i(1)$ and the standardized summand by at most $\{\phi_i(T)-\phi_i(1)\}/\min_sv_i(s)$, which is at most $\sqrt{12}$ for Wilcoxon scores. Thus, we work with rank statistics in the rest of this section. Figure S1 of the Supplement shows the bound matrices, the flipped cells, the reference values under each bound matrix, and the lower bound \eqref{eq:linear-burn} on one row.

The version of the rank statistic to use here is the $c$-adjusted statistic $\psi_R^{c}$, i.e., $\psi_R^{\bfdelta}$ in Example \ref{ex:ranks} with the constant bound $c$. Under the bound matrix $\bfdelta(\calA)$, the imputed row at $\bfs'$ has entries $Y_{it}+c\{W_{it}(s_i')-W_{it}(S_i)\}$ at the cells outside $\calA$. Subtracting $cW_{it}(s_i')$ leaves $Y_{it}-cW_{it}(S_i)$, which does not depend on $\bfs'$. So the reference statistic at every counterfactual onset is computed from one adjusted row, the observed row with $c$ removed from the realized post-onset block, with the flipped cells of $\calA$ sent to $\pm\infty$.

\subsection{Computable solutions}\label{sec:compute}

Fix an $i\in \calI$ and a counterfactual onset $s'$. Let $\calB_i=\{i\}\times\{1,\ldots,T\}$ be the row of unit $i$, and let $\widecheck\bfY_i=\bfY_i-c\bfW_i(S_i)$ be the $c$-adjusted observed row, where $\bfY_i$ and $\bfW_i(S_i)$ are row $i$ of $\bfY$ and of $\bfW(\bfS)$. For a cell set $\calA_i\subseteq\calB_i$, write $\widecheck\Psi_i(s';\calA_i)$ for the score sum $\Psi_i(s';\cdot)$ evaluated at $\widecheck\bfY_i$, with the flipped cells of $\calA_i$ sent to $-\infty$ when untreated under $s'$ and to $+\infty$ when treated under $s'$. This is the unit-$i$ score sum of $\psi_R^{c}$ at the matrix $\widetilde\bfY(\bfs')$ imputed under $\bfdelta(\calA)$, for any $\calA$ with $\calA\cap\calB_i=\calA_i$. The unit-$i$ summand is $\{\widecheck\Psi_i(s';\calA_i)-m_i(s')\}/v_i(s')$. Only $\calA\cap\calF$ affects $p_{\bfS,\bfdelta(\calA)}$, and the $p$-value does not decrease when $\calA\cap\calF$ is enriched. So, when $B-k<|\calB\cap\calF|$, the maximum in \eqref{eq:max-A} is attained at a set $\calA\subseteq\calB\cap\calF$ (Section S2.3 of the Supplement). The exact $p$-value is thus a maximum over $\binom{|\calB\cap\calF|}{B-k}$ sets, each requiring one pass over $\Omega$, which is feasible when $B-k$ is small. Otherwise, we need to use an envelope for a conservative approximation.

\begin{definition}[Envelopes]\label{def:envelope}
    For $\ell\in\{0,\ldots,T\}$, the pointwise envelope is $\overline\Psi_i(s';\ell)=\max\{\widecheck\Psi_i(s';\calA_i):\calA_i\subseteq\calB_i,\ |\calA_i|=\ell\}$. For the Wilcoxon scores $\phi_i(r)=r$, let $R_{it}(\widecheck\bfY_i)$, $t\in\{1,\ldots,T\}$, be the ranks of the $c$-adjusted row $\widecheck\bfY_i$, and define the per-cell gain of $(i,t)\in\calB_i$ at $s'$ by
    \begin{equation}\label{eq:gain}
        g_{it}(s')=\begin{cases}\#\{t':\ W_{it'}(s')=1,\ R_{it'}(\widecheck\bfY_i)<R_{it}(\widecheck\bfY_i)\},& W_{it}(S_i)=1,\ W_{it}(s')=0,\\ \#\{t':\ W_{it'}(s')=0,\ R_{it'}(\widecheck\bfY_i)>R_{it}(\widecheck\bfY_i)\},& W_{it}(S_i)=0,\ W_{it}(s')=1,\\ 0,& \text{otherwise},\end{cases}
    \end{equation}
    and let $\Gamma_i(s';\ell)$ be the sum of the $\ell$ largest values of $g_{it}(s')$ over $t$. For $\ell>T$, set $\overline\Psi_i(s';\ell)=\overline\Psi_i(s';T)$ and $\Gamma_i(s';\ell)=\Gamma_i(s';T)$.
\end{definition}

The envelope $\overline\Psi_i(s';\ell)$ is the largest score sum that unit $i$ can reach at onset $s'$ with $\ell$ of its cells unbounded. The gain $g_{it}(s')$ is the increase in the score sum of unit $i$ at $s'$ when cell $(i,t)$ is the only cell left unbounded, so that it is imputed as $+\infty$ if it is treated under $s'$ and $-\infty$ if not. A cell sent to $-\infty$ drops to the bottom rank, and every cell below it that is treated under $s'$ moves up one rank. Similarly, a cell sent to $+\infty$ rises to the top rank, and the score sum gains one for every cell above it that is untreated under $s'$. Any cell not flipped by $s'$ has zero gain, and $\Gamma_i(s';\ell)$ is the total of the $\ell$ largest gains. Figure \ref{fig:envelope} illustrates the gains and the envelope on one row, and Table S5 of the Supplement gives the numbers.

\begin{figure}[t]
    \centering
    \includegraphics[width=\textwidth]{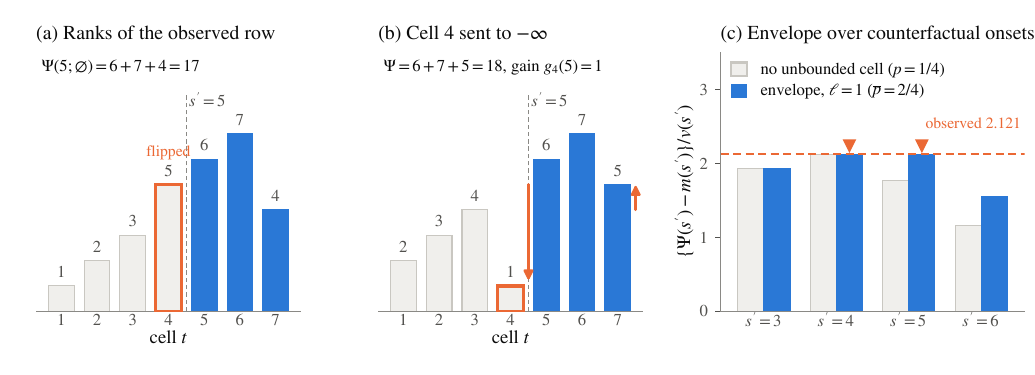}
    \caption{The envelope in Definition \ref{def:envelope} on one unit with $T=7$, observed row $\bfY=(1,2.5,3,5,6.5,7,4)$, realized onset $S=4$, admissible onsets $\{3,4,5,6\}$, Wilcoxon scores, and $c=0$. (a) The ranks under the counterfactual onset $s'=5$. Blue cells are treated under $s'$, and cell $4$ (orange outline) is flipped. (b) Cell $4$ sent to $-\infty$ drops to the bottom rank; cell $7$, the only cell below it that is treated under $s'$, moves up one rank, and the score sum gains $g_4(5)=1$. (c) The standardized rank sum at each admissible onset with no unbounded cell (gray) and with the envelope for one unbounded cell (blue), with the observed value $2.121$ marked. The envelope reaches it at two of the four onsets (triangles), so the envelope $p$-value in Proposition \ref{prop:envelope} is $2/4$, compared with $1/4$ with no unbounded cell.}\label{fig:envelope}
\end{figure}

\begin{proposition}[Envelope $p$-values]\label{prop:envelope}
    Let the conditions of Theorem \ref{thm:quantile} hold, with $\psi$ the sum over $i\in \calI$ of the standardized summands of $\psi_R^{c}$, for any nondecreasing and nonconstant scores. Define
    \begin{equation}\label{eq:envelope-p}
        \overline p_{\bfS,k,c}=\max_{\sum_{i\in \calI}\ell_i=B-k}\ \bbP_{\bfs'}\left\{\sum_{i\in \calI}\frac{\overline\Psi_i(s_i';\ell_i)-m_i(s_i')}{v_i(s_i')}\ \ge\ \psi(\bfS,\bfY)\right\},
    \end{equation}
    i.e., the maximum being over nonnegative integer allocations with $\ell_i\le T$. Then, $\overline p_{\bfS,k,c}\ge p_{\bfS,k,c}$, so that $\bbP(\overline p_{\bfS,k,c}\le\alpha)\le\alpha$ under $H_{k,c}^{\calB}$. For Wilcoxon scores, $\overline\Psi_i(s';\ell)=\widecheck\Psi_i(s';\emptyset)+\Gamma_i(s';\ell)$ for all $(i,s',\ell)$.
\end{proposition}

The proof is in Section S2.4 of the Supplement. For general scores, the pointwise envelope is computed by enumerating the $\binom{T}{\ell}$ subsets of one row. For Wilcoxon scores, the cells a counterfactual onset flips lie on one side of the realized onset and are sent to the same infinity. Because the scores are linear in the rank, the increase in the score sum is then the sum of their single-cell gains. The maximum in \eqref{eq:envelope-p} runs over at most $\binom{B-k+|\calI|-1}{|\calI|-1}$ allocations, which is polynomial in $B-k$ for a fixed $|\calI|$.

The envelope is conservative because it lets the maximizing set vary with the counterfactual onset, while the supremum in \eqref{eq:max-A} fixes one set for all onsets. Proposition S1 of the Supplement gives a sufficient condition under which the two coincide, namely that an allocation attaining \eqref{eq:envelope-p} is realized by one cell set per unit at every onset. For Wilcoxon scores, this holds in particular when $B-k=1$ and, for the unit that receives the unbounded cell, one cell has the largest gain at every admissible onset. Because $\overline\Psi_i(s';\ell)$ is nondecreasing in $\ell$, the same proposition also gives an allocation-free bound,
\begin{equation}\label{eq:allot-free-p}
    \overline p_{\bfS,k,c}\le\bbP_{\bfs'}\left\{\sum_{i\in \calI}\frac{\overline\Psi_i(s_i';B-k)-m_i(s_i')}{v_i(s_i')}\ge\psi(\bfS,\bfY)\right\},
\end{equation}
which is obtained by allotting all $B-k$ cells to every unit at once and requires no search over allocations. Section S5.2 of the Supplement compares the envelope and this bound with the exact supremum by simulation.

\subsection{Simultaneous confidence sets}\label{sec:simul}

Let $\calC_{\alpha}$ be the set of effect matrices not rejected by any quantile test $H_{k,c}^{\calB}$ at level $\alpha$, over all $k$ and $c$. Let $\calC_{\alpha}(c)$ be the set not rejected by any quantile test with the fixed threshold $c$, which constrains only the number of cells of $\calB$ with effects above $c$. Under the conditions of Theorem \ref{thm:quantile}, Theorem S1 of the Supplement gives two simultaneous statements. If the statistic is the same for every $(k,c)$, then $\bbP(\bftau\in\calC_\alpha)\ge1-\alpha$, because a true quantile null is rejected only on the event $\{p_{\bfS,\bftau}\le\alpha\}$. Under the $c$-adjusted rank statistic, $\bbP\{\bftau\in\calC_\alpha(c)\}\ge1-\alpha$ for every fixed $c$, with $p_{\bfS,k,c}$ or with the envelope $\overline p_{\bfS,k,c}$. The $c$-adjusted statistic changes with the threshold, so what remains simultaneous is the family of statements, over $k$, about the number of cells with effects above a given $c$.

\begin{remark}[Relation to single-period quantile inference]\label{rem:caughey-quantile}
    With exchangeable assignment, such as complete randomization, and a rank score statistic, \citet[Theorem 3]{Caughey2023} show that the supremum of their $p$-value over the null is attained by placing the unbounded effects on the treated units with the largest observed outcomes. Their statistic is distribution-free, i.e., its reference distribution does not depend on the outcomes. The supremum is therefore attained where the observed statistic is smallest. Under onset randomization, a counterfactual onset moves a contiguous block of a unit's cells, and the reference distribution of a rank statistic depends on the outcomes and on which cells are unbounded. The cells that raise the reference statistic most can also differ across $\bfs'$ (Section \ref{sec:sup}). Their placement rule therefore need not attain the supremum in \eqref{eq:max-A}, and this holds also when the statistic is evaluated at the imputed control outcomes, as in their work. For stratified experiments, \citet{Su2024} reduce the supremum to a knapsack problem, because the observed statistic is a sum of stratum terms and the statistic is distribution-free. The maximum in \eqref{eq:envelope-p} is instead a probability over onsets of a sum across units, which does not split into unit terms.
\end{remark}

\section{Power with few units and many periods}\label{sec:power}

In this section, we study how the $p$-value of the bounded-null test behaves against a fixed alternative, for the statistics in Examples \ref{ex:stats}(a) and \ref{ex:ranks} under the designs in Example \ref{ex:designs}, in the regime of $N$ fixed and $T\to\infty$. \citet{Liu2026} derived power approximations for switchback experiments (treatment is not absorbing), and the nulls are partially sharp or weak. Throughout this section, Assumption \ref{ass:static} holds, the bound matrix $\bfdelta$ has finite entries, and the alternative has constant excess, i.e., $\tau_{it}=\delta_{it}+\Delta$ with $\Delta>0$. For the mean contrast, the bound is also constant, i.e., $\delta_{it}=\delta$, because the imputed bound enters the contrast through its average over the post-onset block, and a bound that varies over time would act on the reference contrasts as a trend in the baseline. The adjusted rank statistic, on the other hand, removes the bound before ranking and thus needs no such restriction.

\subsection{Designs and baselines}\label{sec:power-setting}

We give the additional assumptions on designs and baseline potential outcomes.

\begin{assumption}[Onset window]\label{ass:window}
    For each $T$, there is a window $\calS_T=\{a_T+1,\ldots,a_T+K_T\}$ with $a_T\ge\rho T$ and $T-a_T-K_T+1\ge\rho T$ for a fixed $\rho\in(0,1/2)$, and the design is uniform on a set $\Omega_T\subseteq\calS_T^N$ with $|\Omega_T|\ge\beta K_T^N$ for a fixed $\beta\in(0,1]$.
\end{assumption}

Assumption \ref{ass:window} places the window inside the series with at least $\rho T$ periods on either side, so every admissible onset leaves long pre-onset and post-onset blocks. The design puts equal mass on a set $\Omega_T$ that is at least a fraction $\beta$ of the full window $\calS_T^N$. Independent uniform onsets have $\Omega_T=\calS_T^N$ and $\beta=1$. Permuted onsets have $|\Omega_T|=N!$, so no fixed $\beta$ satisfies Assumption \ref{ass:window} as $K_T$ grows.

\begin{assumption}[Baseline regularity for means]\label{ass:reg}
    There are constants $\sigma_1,\ldots,\sigma_N>0$, centers $\mu_1,\ldots,\mu_N$, and a sequence $L_T\ge1$ such that for every unit $i$ and every $1\le t_1\le t_2\le T$, $|\sum_{t=t_1}^{t_2}(Y_{it}^{(0)}-\mu_i)|\ \le\ \sigma_i L_T\sqrt{t_2-t_1+1}$.
\end{assumption}

Assumption \ref{ass:reg} states that the baseline of each unit is noise around the center $\mu_i$. The sum of its deviations from $\mu_i$ over any segment of $n$ consecutive periods is at most $\sigma_iL_T\sqrt n$. By a union bound over the at most $NT^2$ segments, a realization of independent sub-Gaussian noise with variance proxy $\sigma_i^2$ satisfies the assumption with $L_T=\{2\log(2NT^2/\theta)\}^{1/2}$ with probability at least $1-\theta$. We write $\sigma_{\max}=\max_i\sigma_i$. The regularity condition for the rank statistic concerns pairs, since it only sees which of two values is larger. Consider a row $\bfy_i$ with distinct entries, two disjoint segments $J$ before $J'$ of lengths $n_J$ and $n_{J'}$, and a shift $x\ge0$. Let $U_{i,J,J'}(x)=\#\{(t,t')\in J\times J':\ y_{it}<y_{it'}+x\}$ be the number of pairs of an earlier period in $J$ and a later period in $J'$ in which the later value, shifted by $x$, is larger.

\begin{assumption}[Baseline regularity for ranks]\label{ass:rank-reg}
    The baseline row of every unit has distinct entries and no two entries differ by exactly $\Delta$, so that the row $\bfY_i^{(0)}+\Delta\bfW_i(S_i)$ also has distinct entries, and there are $\kappa\in(0,1/2]$ and a sequence $L^R_T\ge1$ such that for every unit $i$ and every pair of disjoint segments $J$ before $J'$,
    \begin{align}
        \left|U_{i,J,J'}(0)-\frac12n_Jn_{J'}\right|&\ \le\ L^R_Tn_Jn_{J'}/\sqrt{\min(n_J,n_{J'})},\label{eq:rank-reg-1}\\
        \big|U_{i,J,J'}(\Delta)-U_{i,J,J'}(0)-\kappa n_Jn_{J'}\big|&\ \le\ L^R_Tn_Jn_{J'}/\sqrt{\min(n_J,n_{J'})},\label{eq:rank-reg-2}
    \end{align}
    where $U_{i,J,J'}$ is evaluated at the baseline row $\bfY_i^{(0)}$.
\end{assumption}

Assumption \ref{ass:rank-reg} rules out ties, so that ranks are well-defined before and after the shift. By \eqref{eq:rank-reg-1}, the later value is larger in about half the pairs of any two baseline segments. By \eqref{eq:rank-reg-2}, adding $\Delta$ to the later segment makes it larger in a further fraction $\kappa$ of the pairs, namely those in which the later value was below the earlier one by less than $\Delta$. Thus $\kappa\in(0,1/2]$ measures how visible an effect of size $\Delta$ is to a rank statistic, as $\Delta/\sigma_i$ does in Assumption \ref{ass:reg}. By Hoeffding's inequality for U-statistics \citep{Hoeffding1963}, a realization of independent continuous noise with distribution function $F$ satisfies the assumption with $\kappa=\int\{F(y)-F(y-\Delta)\}\,\rmd F(y)$ and $L^R_T=\{\log(4NT^4/\theta)/2\}^{1/2}$ with probability at least $1-\theta$. For baselines with trends or level shifts, Assumptions \ref{ass:reg} and \ref{ass:rank-reg} hold only with $L_T$ and $L^R_T$ at least of order $T^{1/2}$, and the bounds of Theorems \ref{thm:power-upper} and \ref{thm:power-rank} below then exceed one unless the trend is small relative to $\Delta$, since a trend that mimics the onset is indistinguishable from an effect.

\subsection{The mean contrast}\label{sec:power-mean}

\begin{theorem}[Upper bound, mean contrast]\label{thm:power-upper}
    Let Assumptions \ref{ass:static}--\ref{ass:reg} hold, $\psi=\psi_M$, and $\delta_{it}=\delta$ be constant. Then, for every $\delta\in\bbR$, $\Delta>0$, and realization of $\bfS$, $p_{\bfS,\bfdelta} \le (2\Lambda_T+1)^{N}/(\beta K_T^N)$, with $\Lambda_T=C_1N\rho^{-3}(\sigma_{\max}L_T/\Delta)^2$, where $C_1>0$ is a universal constant. Consequently, the test of $H_{\le\bfdelta}$ at level $\alpha$ rejects with probability one whenever $K_T\ge(\beta\alpha)^{-1/N}(2\Lambda_T+1)$.
\end{theorem}

\begin{figure}[t]
    \centering
    \includegraphics[width=\textwidth]{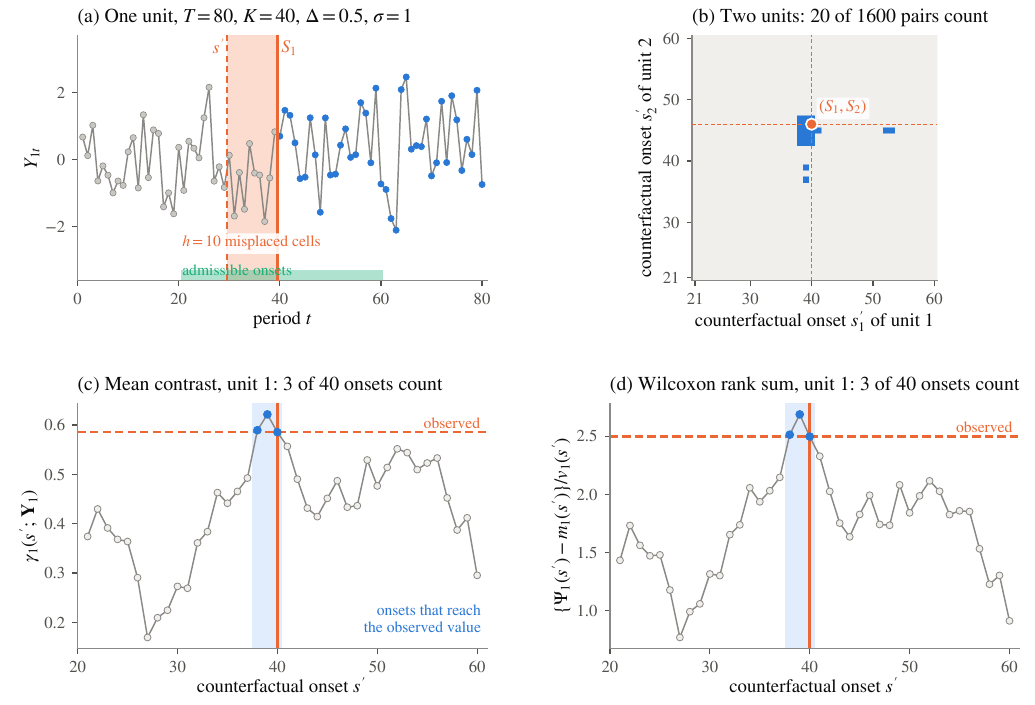}
    \caption{Onsets that reach the observed statistic in one simulated experiment, with $N=2$ units, $T=80$ periods, independent onsets uniform on the window $\{21,\ldots,60\}$ of $K=40$ onsets, realized onsets $(S_1,S_2)=(40,46)$, Gaussian white-noise baselines with $\sigma=1$, constant excess $\Delta=0.5$, sharp null. (a) The series of unit $1$, with the realized onset $S_1$ and a counterfactual onset $s'=30$ that puts $h=10$ cells on the wrong side of the split (shaded). (b) The $40\times40$ onset pairs $(s_1',s_2')$. The pairs at which the mean contrast reaches its observed value (blue) count toward the $p$-value, $20/1600$, and most of them lie in a box around $(S_1,S_2)$. (c) The mean contrast of unit $1$ alone at every counterfactual onset, with its observed value marked. The onsets that reach it (blue) count, $3$ of $40$, and lie within a few periods of $S_1$. (d) The same for the standardized Wilcoxon rank sum in Example \ref{ex:ranks}.}\label{fig:box}
\end{figure}

The proof is in Section S3.1 of the Supplement. A counterfactual onset $s'$ that is $h=|s'-S_i|$ periods away from the true onset of unit $i$ puts $h$ cells on the wrong side of the split. The mean contrast at $s'$ then loses the signal those cells carry, of order $\Delta h/T$, and gains at most the baseline noise of a segment of $h$ cells, of order $\sigma_iL_T\sqrt h/T$ by Assumption \ref{ass:reg}. The noise can cover the lost signal only when $h$ is at most about $(\sigma_iL_T/\Delta)^2$, so the onsets that the test cannot tell from the true one lie within $\Lambda_T$ periods of it. This half-width depends on $T$ only through $L_T$, which is of order $(\log T)^{1/2}$ for sub-Gaussian baselines. The $p$-value is at most the fraction of the design's onset vectors that lie in this box around the truth, that is, at most $(2\Lambda_T+1)^N$ vectors of probability at most $1/(\beta K_T^N)$ each. Thus, the bound improves with the window $K_T$, one factor per unit, but not with the series length. Figure \ref{fig:box} illustrates this in one simulated experiment, in which most onset vectors that reach the observed statistic lie near the realized one.

\begin{proposition}[Lower bounds, mean contrast]\label{prop:power-lower}
    (a) For every design, statistic, $\bfdelta$, and realization, $p_{\bfS,\bfdelta}\ge\bbP_{\bfs'}(\bfs'=\bfS)\ge\min_{\bfs\in\Omega}\bbP(\bfS=\bfs)$, which is $1/|\Omega_T|\ge K_T^{-N}$ under Assumption \ref{ass:window} and $1/N!$ for every $T$ under the permuted-onset design.
    (b) Let Assumptions \ref{ass:static} and \ref{ass:window} hold with $\beta=1$, $\psi=\psi_M$, $\delta_{it}=\delta$ be constant, and $\tau_{it}=\delta+\Delta$ with $\Delta>0$. Let $\sigma>0$ and $h_0=\lfloor C_2(\sigma/\Delta)^2\rfloor$, where $C_2>0$ is a universal constant, and suppose that $h_0\ge1$ and $K_T\ge2h_0$. There exist baseline sequences satisfying Assumption \ref{ass:reg} with $\sigma_i=\sigma$ and $L_T=\{2\log(NT^{2N+4})\}^{1/2}$ for which $\bbE(p_{\bfS,\bfdelta}) \ge (C_3h_0/K_T)^N$ for all $T\ge T_0$, where $C_3>0$ is a universal constant and $T_0=T_0(\rho,N,\sigma/\Delta)$ is a threshold depending only on its arguments, the expectation being over the design.
\end{proposition}

Part (a) holds because the realized onset vector is always in the reference set and reproduces the observed statistic, so the $p$-value is at least its probability. Part (b) is proved in Section S3.4 of the Supplement. For some baselines allowed by Assumption \ref{ass:reg}, the noise in the $h_0$ cells just after the onset is as large as the signal those cells carry, so a constant fraction of the onsets up to $h_0$ periods away reach the observed statistic. When $h_0\ge1$, $h_0$ is of the order of $(\sigma/\Delta)^2$ and $\Lambda_T$ of the order of $(\sigma L_T/\Delta)^2$, so the lower bound matches the upper bound of Theorem \ref{thm:power-upper} up to the factor $L_T^{2N}$ and factors depending only on $N$ and $\rho$. For larger effects, $h_0=0$ and part (a) gives the floor $K_T^{-N}$, while $\Lambda_T$ is of the order of $L_T^2$, so the rate $K_T^{-N}$ holds for every effect size. Section S3.2 of the Supplement extends Theorem \ref{thm:power-upper} to an excess that varies with time within a factor $(1+\rho)$ of its smallest value.

\subsection{The Wilcoxon rank sum}\label{sec:power-rank}

This subsection gives the rank counterpart of Section \ref{sec:power-mean}, with the Wilcoxon rank sum in Example \ref{ex:ranks}. For Wilcoxon scores, the score sum of the post-onset block equals, up to its centering, the number of pre--post pairs where the post value is larger, which is controlled by Assumption \ref{ass:rank-reg}. The test uses the $\bfdelta$-adjusted statistic $\psi_R^{\bfdelta}$, which removes the bound before ranking, so every counterfactual onset is evaluated on the same row (Remark S2 of the Supplement).

\begin{theorem}[Upper bound, rank statistic]\label{thm:power-rank}
    Let Assumptions \ref{ass:static}, \ref{ass:window}, and \ref{ass:rank-reg} hold, $\psi=\psi_R^{\bfdelta}$ with Wilcoxon scores, and $\tau_{it}=\delta_{it}+\Delta$. Then, for every realization of $\bfS$, $p_{\bfS,\bfdelta} \le (2\Lambda^R_T+1)^{N}/(\beta K_T^N)$, with $\Lambda^R_T=C_4N\rho^{-8}(L^R_T/\kappa)^2$, where $C_4>0$ is a universal constant. Consequently, the $\bfdelta$-adjusted rank test of $H_{\le\bfdelta}$ at level $\alpha$ rejects with probability one whenever $K_T\ge(\beta\alpha)^{-1/N}(2\Lambda^R_T+1)$.
\end{theorem}

The proof is in Section S3.3 of the Supplement. The argument has the same form as for Theorem \ref{thm:power-upper}. A counterfactual onset $h$ periods away moves $h$ cells across the split. Under the true onset, the excess has reversed the order of a fraction $\kappa$ of the pairs these cells form with the cells on the other side of $S_i$. Under $s'$, these pairs are no longer counted. Measured from its centering and standardized, the count therefore loses a signal of order $\kappa h/\sqrt T$ and gains at most a two-sample fluctuation of order $L^R_T\sqrt{h/T}$, by Assumption \ref{ass:rank-reg}. The signal is proportional to $\kappa$ rather than $\Delta$ because a rank statistic sees the excess only through the fraction of pairs whose order it reverses, which cannot exceed $1/2$ regardless of how large $\Delta$ is. The onsets the test cannot tell from the true one then lie within $\Lambda^R_T$ periods of it, and the $p$-value is at most the share of onset vectors in that box.

\begin{proposition}[Lower bounds, rank statistic]\label{prop:rank-lower}
    (a) For every design, statistic, $\bfdelta$, and realization, $p_{\bfS,\bfdelta}\ge\bbP_{\bfs'}(\bfs'=\bfS)\ge\min_{\bfs\in\Omega}\bbP(\bfS=\bfs)$, which is $1/|\Omega_T|\ge K_T^{-N}$ under Assumption \ref{ass:window} and $1/N!$ for every $T$ under the permuted-onset design. (b) Let Assumptions \ref{ass:static} and \ref{ass:window} hold with $\beta=1$, $\psi=\psi_R^{\bfdelta}$ with Wilcoxon scores, and $\tau_{it}=\delta_{it}+\Delta$ with $\Delta>0$, and let $F$ be a continuous distribution function with a bounded density $f$. Put $\overline\kappa=\Delta\sup f$, $\kappa'=(1+C_5\rho^{-3})\overline\kappa$, and $h_0^R=\lfloor C_6/\kappa'^2\rfloor$, where $C_5,C_6>0$ are universal constants, and suppose that $h_0^R\ge1$ and $K_T\ge2h_0^R$. There exist baseline sequences satisfying Assumption \ref{ass:rank-reg} with $\kappa=\int\{F(y)-F(y-\Delta)\}\,\rmd F(y)$ and $L^R_T=\{\log(4NT^{2N+6})/2\}^{1/2}$ for which $\bbE(p_{\bfS,\bfdelta}) \ge (C_7h_0^R/K_T)^N$ for all $T\ge T_0^R$, where $C_7>0$ is a universal constant and $T_0^R=T_0^R(\rho,N,\kappa')$ is a threshold depending only on its arguments, the expectation being over the design.
\end{proposition}

Part (a) holds for every statistic, as in Proposition \ref{prop:power-lower}(a). Part (b) is proved in Section S3.5 of the Supplement. For some baselines allowed by Assumption \ref{ass:rank-reg}, the $h_0^R$ cells just before the onset sit high enough in the baseline distribution that moving them across the split changes the count of pairs by as much as the excess does. A constant fraction of the onsets up to $h_0^R$ periods away then reach the observed statistic, and $h_0^R$ is of the order of $(\kappa')^{-2}$. When $h_0^R\ge1$, the lower bound therefore matches Theorem \ref{thm:power-rank} in its dependence on $\kappa^{-2N}K_T^{-N}$, up to $(L^R_T)^{2N}$, the factor $(\kappa'/\kappa)^{2N}$, and factors depending only on $N$ and $\rho$. For Gaussian baselines, $\kappa'/\kappa$ is at most a constant depending on $\rho$ throughout the range $h_0^R\ge1$ where part (b) applies (Section S3.6 of the Supplement). For larger effects, $h_0^R=0$, and the rate $K_T^{-N}$ still holds up to the same factors. Theorem S2 in Section S3.7 of the Supplement combines these upper and lower bounds into bounds on the worst-case $p$-value over the baseline classes of Assumptions \ref{ass:reg} and \ref{ass:rank-reg}.

\section{Planning an experiment}\label{sec:planning}

The results of Section \ref{sec:power} inform the choices made before an experiment starts, namely how many units to enroll, how many periods to run, how wide to make the onset window, and which design to use. Section \ref{sec:planning-rules} states the rules that follow from the theory, and Section \ref{sec:planning-sim} calibrates the window needed for a target power by simulation. We write $K$ for the number of admissible onsets of each unit.

\subsection{Rules from the theory}\label{sec:planning-rules}

\emph{Feasibility.} By Proposition \ref{prop:power-lower}(a), the $p$-value under a uniform design is at least $1/|\Omega|$, so no test can reject at level $\alpha$ unless $|\Omega|\ge1/\alpha$. With independent uniform onsets, $N$ units need $K^N\ge1/\alpha$, and one unit needs at least $20$ admissible onsets at level $0.05$. Under the permuted-onset design, $|\Omega|=N!$, so at least four units are needed at level $0.05$. For one unit, the test at level $\alpha$ rejects only when at most $\lfloor\alpha K\rfloor$ admissible onsets reach the observed statistic, so a window whose size is a multiple of $1/\alpha$ makes $\alpha$ an attainable value of the $p$-value.

\emph{Periods and the window.} Theorems \ref{thm:power-upper} and \ref{thm:power-rank} bound the $p$-value through a box whose half-width depends on $T$ only through $L_T$ and $L^R_T$. For some baselines, Propositions \ref{prop:power-lower}(b) and \ref{prop:rank-lower}(b) bound the expected $p$-value from below by a quantity that depends on the window, not on $T$. A longer series therefore raises power only by allowing a wider window. Section S5.5 of the Supplement holds the window at $40$ onsets and lets $T$ grow from $80$ to $640$, and the power does not increase. Section S5.6 holds $T$ fixed and widens the window until its first onset has only five periods before it and its last onset only five periods after it. The power of the rank sum then keeps rising. The mean contrast does best when the first onset has about $15$ periods before it and the last about $15$ after it, because at an onset near either end of the window it averages only a few values on one side of the split. When the window comes close to the ends of the series, the rank sum is the statistic to plan with.

\emph{Scaling with the effect.} At a fixed $\rho$, the window that Theorems \ref{thm:power-upper} and \ref{thm:power-rank} show to be sufficient scales with $(\sigma/\Delta)^2$ for the mean contrast and with $\kappa^{-2}$ for the rank sum. The constants $C_1$ and $C_4$ are far from sharp, so these windows are much wider than needed, and only how they scale is used here. When $h_0\ge1$ and $h_0^R\ge1$, the lower bounds of Propositions \ref{prop:power-lower}(b) and \ref{prop:rank-lower}(b) match these scalings up to logarithmic factors, factors depending only on $N$ and $\rho$, and, for the rank sum, the factor $(\kappa'/\kappa)^{2}$. For Gaussian baselines, this factor is at most a constant depending on $\rho$ (Theorem S2 and Section S3.6 of the Supplement). A test with power $\pi$ has $\bbE(p_{\bfS,\bfdelta})\le1-\pi(1-\alpha)$, so these lower bounds also limit the power at a given window. At a fixed number of units, level, and $\rho$, the window needed for a given power should thus scale in the same way.

\emph{Adding units or periods.} With independent onsets, each additional unit multiplies the number of onset vectors by its own number of admissible onsets. Table S15 of the Supplement shows how much this lowers the window each unit needs for the same power. The total number of periods, summed over the units, need not fall, because every unit is observed over its own series.

\emph{Choice of design.} The upper bounds depend on the design only through $\beta$, and the window they show to be sufficient, $(\beta\alpha)^{-1/N}(2\Lambda_T+1)$ or $(\beta\alpha)^{-1/N}(2\Lambda^R_T+1)$, grows by the factor $\beta^{-1/N}$ when the design is restricted. For the staggered windows of \citet{Koehler1998} with $N$ dividing $K$, Proposition S2 of the Supplement gives $\beta=N!/N^N$, so the factor is below $e$ and at most $1.81$ for $N\le4$. For the minimum spacing of \citet{Levin2018}, with $\underline h$ periods between onsets, the factor is at most $\{1-(N-1)\underline h/K\}^{-1}$. Section S3.8 of the Supplement shows that the lower bounds keep their rate under both designs, under staggered windows once $\lfloor K/N\rfloor\ge2h_\ast$ and under minimum spacing once $K\ge(N-1)\underline h+2Nh_\ast$, where $h_\ast=h_0$ for the mean contrast and $h_\ast=h_0^R$ for the rank sum. Permuted onsets keep the $p$-value at or above $1/N!$ whatever the window. Corollary S1 of the Supplement provides these comparisons.

\emph{Serial dependence.} For AR(1) noise with autocorrelation $\lambda$, the sums of baseline deviations over long segments scale with the long-run standard deviation, whose square is $(1+\lambda)/(1-\lambda)$ times the marginal variance, three times at $\lambda=0.5$. With $\sigma_i$ the marginal standard deviation, Assumption \ref{ass:reg} then holds only with $L_T$ larger by the square root of this factor, and the pair counts of Assumption \ref{ass:rank-reg} fluctuate more in the same way. The window needed should thus be about three times the window needed under white noise.

\subsection{Window needed for a target power}\label{sec:planning-sim}

\begin{figure}[t]
    \centering
    \includegraphics[width=\textwidth]{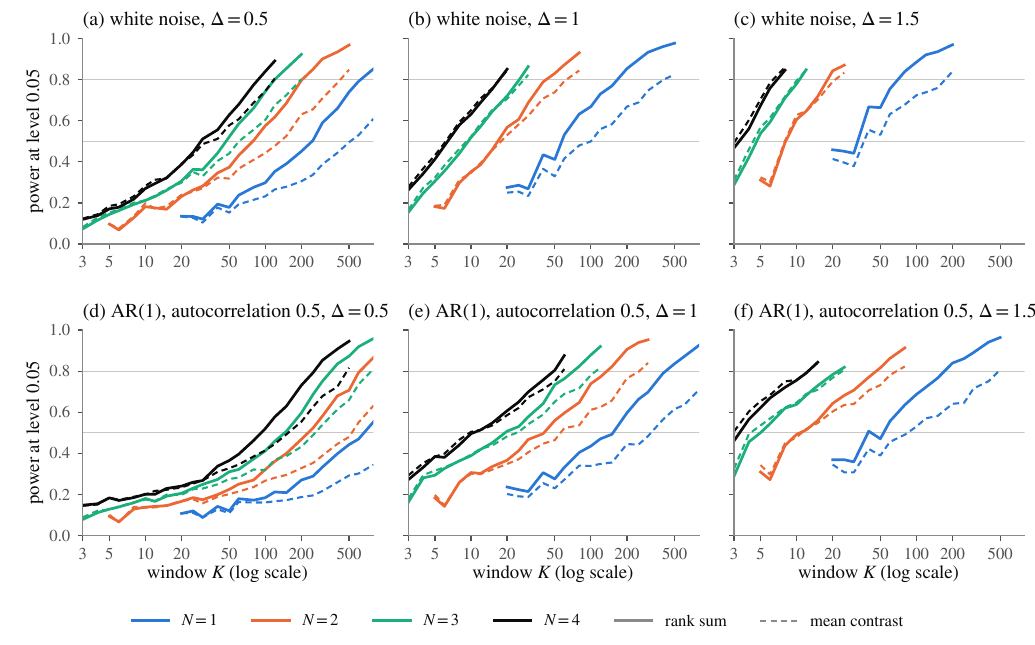}
    \caption{Power at level $0.05$ against the window $K$ (log scale) in the planning study of Section \ref{sec:planning-sim}, with $T=K+9$ periods per unit, independent onsets uniform on $\{6,\ldots,K+5\}$, and $2{,}000$ replications per window. Columns: effect $\Delta\in\{0.5,1,1.5\}$ in every treated cell. Top row: Gaussian white-noise baselines. Bottom row: Gaussian AR(1) baselines with autocorrelation $0.5$. Both baselines have unit variance. Solid lines are the Wilcoxon rank sum, dashed lines the mean contrast, and colors the number of units $N$. The gray lines mark power $0.5$ and $0.8$.}\label{fig:planning}
\end{figure}

This study gives the smallest window at which the tests reach a target power. Each of $N\in\{1,2,3,4\}$ units is observed over $T=K+9$ periods, with independent onsets uniform on the window $\{6,\ldots,K+5\}$ of $K$ onsets. Every admissible onset thus has at least five periods before it and five after it, the minimum numbers of observations before and after the intervention that the single-case design standards of \citet{KratochwillWWC2010} require to meet the standards without reservations. The baselines have unit variance and are Gaussian white noise or Gaussian AR(1) series with autocorrelation $\lambda=0.5$, and every treated cell has the effect $\Delta\in\{0.5,1,1.5\}$. The window runs over a grid from $3$ to $800$ onsets, omitting the windows with $K^N<20$, at which no test can reject at level $0.05$. For each configuration, larger windows are not run once both tests reach power $0.8$. For each configuration and window, the sharp null is tested at level $0.05$ with $\psi_M$ and with the Wilcoxon rank sum in Example \ref{ex:ranks} over $2{,}000$ replications. The $p$-value is exact, computed by splitting the units into two groups and counting the pairs of partial sums whose total reaches the observed statistic. With a constant bound, both the mean contrast and the $\bfdelta$-adjusted rank sum reduce the test of $H_{\le\delta}$ against $\tau_{it}=\delta+\Delta$ to the sharp-null test on $\bfY-\delta\bfW(\bfS)$, so the same power holds for these bounded nulls. Table S15 in Section S5.7 of the Supplement reports the smallest window on the grid at which the power is at least $0.5$ or $0.8$ at that window and at every larger window simulated, and Figure \ref{fig:planning} shows the power curves.

Table S15 of the Supplement shows that one or two units need wide windows unless the effect is large. For the rank sum with white noise and $\Delta=1$, power $0.8$ needs $200$ onsets with one unit, $50$ with two, $30$ with three, and $20$ with four. Additional units shorten the window each unit needs, but not always the total number of periods. In the same configuration, $N(K+9)$ falls from $209$ periods with one unit to $118$ with two and stays between $116$ and $118$ with three and four. The rank sum is the better statistic to plan with. It never needs a wider window than the mean contrast for power $0.8$, and for power $0.5$ it needs a wider one only in three configurations, all with windows of at most $12$ onsets. With one unit and $\Delta=0.5$, the mean contrast does not reach power $0.8$ at $800$ onsets, where the rank sum has power $0.85$.

Halving the effect from $1$ to $0.5$ with white noise requires a window $3.75$ to $5$ times as wide for the rank sum and $5$ to $8$ times as wide for the mean contrast. At a fixed $\rho$, Theorems \ref{thm:power-upper} and \ref{thm:power-rank} predict somewhat smaller factors, $4$ for the mean contrast, the growth of $(\sigma/\Delta)^2$, and about $3.5$ for the rank sum, the growth of $\kappa^{-2}$ for Gaussian baselines, where $\kappa=\Phi\{\Delta/(\sigma\sqrt2)\}-1/2$ equals $0.26$ at $\Delta=1$ and $0.14$ at $\Delta=0.5$. In this study, the first onset has five periods before it and the last onset five periods after, so $\rho=5/(K+9)$ in Assumption \ref{ass:window}, and $\rho$ falls as $K$ grows. The sufficient windows in Theorems \ref{thm:power-upper} and \ref{thm:power-rank} widen as $\rho$ falls, and the mean contrast also loses power at onsets with few periods before or after them (Section S5.6 of the Supplement). Autocorrelation $0.5$ widens the window by about a factor of $3$ of the long-run variance. Whenever both windows are on the grid, and the white-noise window has at least $20$ onsets, the window for power $0.8$ is $2.5$ to $3.3$ times as wide for the rank sum. For the mean contrast, it is $2.5$ to $4.2$ times as wide. 

With one or two units, the attainable sizes of the test shape the smallest windows. With one unit, the test at level $0.05$ rejects only when at most $\lfloor0.05K\rfloor$ onsets reach the observed statistic (Section \ref{sec:planning-rules}), so its size is $0.05$ at $20$ onsets but $0.033$ at $30$. In Figure \ref{fig:planning}, the power curves for one unit therefore stay flat or fall, up to Monte Carlo error, from $20$ to $30$ onsets and rise again at $40$. Two units show the same pattern at the smallest windows. With white noise and $\Delta=1.5$, both tests need $40$ onsets for power $0.5$, twice the $20$ at which rejection first becomes possible. Consecutive windows on the grid differ by $20$ to $33$ percent, which limits the precision of all these ratios.

\section{Application to a multiple-baseline experiment}\label{sec:data}

\citet{Laski1988} trained parents of eight children with autism to use the natural language paradigm and scored ten-minute periods of play by partial-interval recording, 60 ten-second intervals per period, the outcome being the percentage of intervals with a child vocalization. The data are distributed with the \texttt{R} package \texttt{scdhlm} \citep{Pustejovsky2026}. The children are observed over $T_i=13,13,16,17,16,14,19,20$ periods with onsets at periods $5,6,8,11,8,8,9,12$, for $B=\sum_iT_i=128$ cells. The original study staggered the intervention without randomizing the onsets, so the analysis is an illustration under a hypothetical design, taken to be independent onsets uniform on $\calS_i=\{4,\ldots,T_i-3\}$. This window leaves at least three periods before and four after every admissible onset, and it contains every realized onset. The window sizes $K_i$ are $7,7,10,11,10,8,13,14$, and $|\Omega|=\prod_iK_i\approx7.8\times10^7$. The statistics are $\psi_M$ and the $\delta$-adjusted Wilcoxon rank sum $\psi_R^{\delta}$. The alternative is that the intervention increases vocalization, and every $p$-value is exact, i.e., unit-level $p$-values enumerate the unit's window, and pooled $p$-values enumerate the sums over the first four children (cases 1, 2, 4, 5) and over the last four (cases 6 to 9) separately and count the pairs whose total reaches the observed statistic (Section S6.1 of the Supplement).

\emph{Unit-level results.} Table \ref{tab:L1} reports, for each child, the sharp-null $p$-values of the two statistics under the child's own window. It also reports a $90\%$ lower confidence bound for $\max_t\tau_{it}$, the infimum of the $\delta$ at which the mean-contrast test of $H_{\le\delta}$ is not rejected at $\alpha=0.10$. The bound is valid by Theorem \ref{thm:bounded}. By Proposition \ref{prop:mono}, the non-rejected set is an interval, and the bound is its left end. The last two columns give the number of the child's cells established at $\alpha=0.10$ to have effects above $c=0$ and $c=10$, from the exact quantile test \eqref{eq:max-A} on the child's own cells with the $c$-adjusted rank statistic. Cases 2, 5, 6, 7, and 9 attain the minimal $p$-value $1/K_i$ with both statistics, case 1 with the rank statistic, and case 4 with the mean contrast. Case 8 shows no evidence of an effect.

\begin{table}[ht]
    \caption{Unit-level results under independent uniform onsets on $\{4,\ldots,T_i-3\}$. Columns: number of periods, onset, window size, sharp-null $p$-values, $90\%$ lower confidence bound for $\max_t\tau_{it}$ from the mean contrast (none when no $\delta\ge-20$ is rejected at level $0.10$), and the number of cells with effect above $c$ established at level $0.10$ by the exact quantile test with the $c$-adjusted rank statistic. Case labels are those of the data set, which has no case 3.}\label{tab:L1}
    \centering
    \footnotesize
    \begin{tabular}{lrrrrrrrr}
        \toprule
        Case & $T_i$ & $S_i$ & $K_i$ & $p$ (mean) & $p$ (rank) & LCB$_{90}$ for $\max_t\tau_{it}$ & cells $>0$ & cells $>10$ \\
        \midrule
        1 & 13 & 5 & 7 & 0.286 & 0.143 & none & 0 & 0 \\
        2 & 13 & 6 & 7 & 0.143 & 0.143 & none & 0 & 0 \\
        4 & 16 & 8 & 10 & 0.100 & 0.300 & 2.90 & 0 & 0 \\
        5 & 17 & 11 & 11 & 0.091 & 0.091 & 17.03 & 1 & 1 \\
        6 & 16 & 8 & 10 & 0.100 & 0.100 & 28.20 & 1 & 1 \\
        7 & 14 & 8 & 8 & 0.125 & 0.125 & none & 0 & 0 \\
        8 & 19 & 9 & 13 & 0.769 & 0.538 & none & 0 & 0 \\
        9 & 20 & 12 & 14 & 0.071 & 0.071 & 19.51 & 1 & 1 \\
        \bottomrule
    \end{tabular}
\end{table}

\emph{Pooled bounded nulls.} By pooling over the eight children, the data establish a large effect. Both statistics reject the pooled sharp null with $p<10^{-4}$. The $95\%$ lower confidence bound for $\max_{i,t}\tau_{it}$ is $19.77$ percentage points with the mean contrast and $19.04$ with the rank statistic. With $95\%$ confidence, the intervention raised some child's vocalization by at least about $20$ percentage points in some period, where the pre-onset averages are about $38$ percent of intervals. Table \ref{tab:L2} reports the exact pooled $p$-values of $H_{\le\delta}$ on a grid of $\delta$. For the mean contrast, the bound is the infimum of the non-rejected set $\{\delta:p_{\bfS,\delta}>\alpha\}$, which is valid by Theorem \ref{thm:bounded} without assuming how effects vary across children or periods. By Proposition \ref{prop:mono}, the non-rejected set is an interval. For the $\delta$-adjusted rank statistic, the observed value moves with $\delta$, so Proposition \ref{prop:mono} does not apply. The infimum of the non-rejected set is still a valid $(1-\alpha)$-lower bound, because if it exceeded $\delta_0=\max_{i,t}\tau_{it}$, the true null $H_{\le\delta_0}$ would be rejected, which happens with probability at most $\alpha$ by Theorem \ref{thm:bounded}. On these data, this set is also an interval (Section S6.2 of the Supplement). At $\alpha=0.10$, the two bounds are $22.01$ and $21.03$. Both are below the bound of $28.20$ for case $6$ in Table \ref{tab:L1}, because the pooled test applies the bound $\delta$ to every child. For $\delta$ between $22$ and $28$, the observed contrast of case $8$ lies far below its reference values. The unit-level bound is a lower confidence bound for $\max_{i,t}\tau_{it}$ only if the child is chosen before the data are collected.

\begin{table}[ht]
    \caption{Pooled bounded-null $p$-values of $H_{\le\delta}$ for the mean contrast and the $\delta$-adjusted rank statistic, computed exactly over the product design, rounded to two significant digits. The mean-contrast entry at $\delta=22$ is $0.0998$.}\label{tab:L2}
    \centering
    \footnotesize
    \resizebox{\textwidth}{!}{
    \begin{tabular}{lrrrrrrrrrr}
    \toprule
    $\delta$ & 0 & 5 & 10 & 15 & 18 & 19 & 20 & 21 & 22 & 30 \\
    \midrule
    mean contrast & 0.000036 & 0.00020 & 0.0014 & 0.0093 & 0.027 & 0.039 & 0.054 & 0.074 & 0.10 & 0.55 \\
    rank & 0.0000078 & 0.00040 & 0.0032 & 0.023 & 0.039 & 0.045 & 0.077 & 0.088 & 0.13 & 0.79 \\
    \bottomrule
    \end{tabular}
    }
\end{table}

\begin{table}[ht]
    \caption{Quantile inference on all $128$ cells with the $c$-adjusted rank statistic, by the exact supremum in \eqref{eq:max-A}, the envelope in \eqref{eq:envelope-p}, and the allocation-free bound in \eqref{eq:allot-free-p}. Rejection establishes at least $B-k+1$ cells with effects above $c$ points. Entries are rounded to two significant digits.}\label{tab:app1}
    \centering
    \footnotesize
    \begin{tabular}{rrrrr}
        \toprule
        $c$ & $B-k$ & exact & envelope & allocation-free \\
        \midrule
        0 & 0 & 0.0000078 & 0.0000078 & 0.0000078 \\
        0 & 1 & 0.000047 & 0.000050 & 0.0057 \\
        0 & 2 & 0.00011 & 0.00015 & 0.059 \\
        0 & 3 & 0.00024 & 0.00035 & 0.15 \\
        10 & 0 & 0.0032 & 0.0032 & 0.0032 \\
        10 & 1 & 0.0069 & 0.0075 & 0.12 \\
        10 & 2 & 0.012 & 0.016 & 0.47 \\
        10 & 3 & 0.017 & 0.028 & 0.74 \\
        20 & 0 & 0.077 & 0.077 & 0.077 \\
        20 & 1 & 0.12 & 0.12 & 0.64 \\
        20 & 2 & 0.16 & 0.19 & 0.94 \\
        20 & 3 & 0.19 & 0.27 & 0.99 \\
        \bottomrule
    \end{tabular}
\end{table}

\emph{Pooled quantile nulls.} The quantile tests show that the effect is not confined to a few periods. At $\alpha=0.05$, at least $14$ of the $72$ cells that some pair of admissible onsets flips have positive effects, and at least $5$ have effects above $10$ points. At $\alpha=0.10$, these counts rise to $19$ and $7$, and at least one cell has an effect above $20$ points. By Theorem S1 of the Supplement, the statements at a common threshold hold simultaneously across $k$. These counts come from the envelope in \eqref{eq:envelope-p}, with $B-k$ raised one step at a time until the envelope exceeds the level (Section S6.3 of the Supplement). The envelope is nearly as sharp as the exact supremum in \eqref{eq:max-A}. Table \ref{tab:app1} compares the two, together with the allocation-free bound in \eqref{eq:allot-free-p}, for $B-k\le3$ and $c\in\{0,10,20\}$. In every row, the envelope is at most $1.7$ times the exact supremum, and the two lead to the same decision at $\alpha=0.05$ and $0.10$. The allocation-free bound is much looser. At $\alpha=0.05$, it establishes only that at least two cells have positive effects and one has an effect above $10$ points.

\emph{The design and the planning rules.} Under the hypothetical design, a child with $K_i$ admissible onsets has the smallest attainable $p$-value $1/K_i$. Since every $K_i$ is below $20$, no child can reject alone at level $0.05$, regardless of the effect size, and cases 1, 2, and 7, with $K_i<10$, cannot reject alone at level $0.10$. For cases 2, 5, 6, 7, and 9, which reach this floor with both statistics in Table \ref{tab:L1}, the unit-level evidence is limited by the design rather than by the data. Pooling multiplies the windows, and the floor of the pooled design, $1/|\Omega|\approx1.3\times10^{-8}$, is far below any conventional level. A randomized replication in which each child can reject alone at level $0.05$ needs at least $20$ admissible onsets per child, that is, at least $29$ periods per child with five periods before the first onset and after the last, as in Table S15 of the Supplement.

The results above rely on Assumption \ref{ass:static}, which Section S6.4 of the Supplement relaxes. Suppose that a treated cell's outcome may change by up to $5$ points with the duration of treatment, or an untreated cell's outcome by up to $5$ points with the timing of the future onset. The null of no positive effect is then still rejected at level $0.05$ with both statistics, but the $95\%$ lower confidence bound falls from about $20$ points to below $6$.

\section{Discussion}\label{sec:discussion}

This paper studied randomization inference for bounded nulls and quantiles of cell effects in randomized experiments under absorbing onset designs. Under the treatment-duration irrelevance assumption, the outcomes that would have been observed under any other onset can be imputed under a bounded null, and the randomization test built on the imputed tables is valid for every onset design and every effect-increasing statistic. The same idea extends to quantile nulls. The supremum over such a null is a maximum over finitely many cell sets, and an envelope bound makes it computable when the enumeration is too expensive. For the mean contrast and the adjusted Wilcoxon rank sum, the $p$-value is at most the fraction of the design's onset vectors that lie within a distance of the true one. When the baselines are independent noise, sub-Gaussian for the mean contrast and continuous with a common distribution for the rank sum, this distance grows only logarithmically in $T$. The bound therefore improves with the number of admissible onsets rather than with the length of the series. For small effects and independent uniform onsets, the worst case of this bound over the baseline class is attained up to factors depending only on $N$ and $\rho$ and a factor polylogarithmic in $T$. For the rank sum, there is an additional factor that depends only on $\rho$ for Gaussian baselines. For planning, a design needs at least $1/\alpha$ onset vectors before any test can reject; a longer series raises power only through the wider window it admits, and Table S15 of the Supplement gives the window needed for a target power. 
 
Several questions remain. The power theory covers a constant excess. The upper bound for the mean contrast also covers an excess that varies over time within a factor $1+\rho$ of its smallest value. Effects that build up with the duration of treatment call for a different statistic. Conditional randomization tests in the manner of \citet{Shaikh2021} and \citet{Zhang2025}, which condition on part of the realized onsets, may offer another route to onset-dependent potential outcomes.



\singlespacing
\bibliographystyle{jasa3}
\bibliography{absorbing_onset}

\end{document}